\documentclass{emulateapj}
\usepackage{longtable}
\usepackage{epsfig} 
\usepackage{times}
\def\mgii{Mg~{\sc ii}~} 
\def\mgiia{Mg~{\sc ii}$\lambda$2796~} 
 
\def\mgiiab{Mg~{\sc ii}$\lambda\lambda$2796,2803~} 

\def\civ{C~{\sc iv}~}
\def\siiv{Si~{\sc iv}~} 
\def\feii{Fe~{\sc ii}~} 
\def\kms{km.s$^{-1}~$} 

\shorttitle{ \mgii absorption systems towards FSRQs} 
\shortauthors{Hum Chand \& Gopal-Krishna}

\begin{document}
\title{Incidence of \mgii absorption systems towards flat-spectrum radio 
quasars}

\author{Hum Chand\altaffilmark{1} and Gopal-Krishna\altaffilmark{2}}
\affil{
$^1$Aryabhatta Research Institute of Observational Sciences (ARIES),
 Manora Peak, Nainital$-$ 263129, India \\
{\tt hum@aries.res.in (HC), krishna@ncra.tifr.res.in(GK)}}
\affil{
$^2$ NCRA.TIFR, Pune University Campus, Pune$-$411007, India}

\begin{abstract}
The conventional wisdom that the rate of incidence of \mgii absorption
systems, $dN/dz$ (excluding  `associated systems' having velocity 
$\beta$c relative to the AGN of less than $\sim$ 5000 km.s$^{-1}$) is totally 
independent of the background AGN, has been challenged by a recent 
finding that $dN/dz$ for strong \mgii absorption systems towards distant 
blazars is $2.2\pm_{0.6}^{0.8}$ times the value known for normal 
optically-selected 
quasars (QSOs).  This has led to the suggestion that a significant 
fraction of even the absorption systems with $\beta$ as high as $\sim 0.1$ 
may have been ejected by the relativistic jets in the blazars, which are 
expected to be pointed close to our direction. Here we investigate this 
scenario using a large sample of  { 115} flat-spectrum radio-loud quasars 
(FSRQs) which too possess powerful jets, but are only weakly polarized.
We show, for the first time, that $dN/dz$ { towards} FSRQs is, on the whole, 
quite similar to that known for QSOs and the comparative excess of strong 
\mgii absorption systems seen towards blazars is mainly confined to $\beta 
< 0.15$.  The excess relative to FSRQs  can probably result from a 
likely closer alignment of blazar jets with our direction
and  hence any gas clouds accelerated by them are more likely to be on the 
line of sight to the active quasar nucleus. 

\end{abstract}

\section{Introduction}
Study of Mg II absorption line systems in the spectra of QSOs
has provided a means of detecting distant normal field galaxies which 
happen to be situated close to the lines of sight to the QSOs 
(e.g., Bergeron et al. 1991; Steidel et al. 1994). 
Barring the so called `associated systems' (having $z_{abs} \sim z_{QSO}$), 
the absorption line systems are customarily believed to arise in intervening 
structures that are wholly unrelated to the background QSO. This view was,
however, 
challenged by the finding that the occurrence rate, $dN/dz$, of \mgii 
absorption systems in the spectra of GRBs is nearly 4 times the value 
found for QSOs, when strong lines having rest-frame equivalent width 
$> 1 \AA$~ are considered (Prochter et al. 2006a). Later studies have 
supported this unexpected trend, albeit the excess factor is found to be 
smaller, 2.1 $\pm$ 0.6 (Sudilovsky et al.  2007; Vergani et al. 2009; 
Tejos et al. 2009).  More recently, Bergeron, Boisse \& Menard (2011, 
hereinafter BBM) have 
examined this issue for the case of blazars which, like GRBs, also possess 
relativistic jets pointed close to our direction, but are considerably 
less variable. Using intermediate-resolution optical/UV spectra of their sample
of 45 powerful blazars (predominantly distant BL Lacs) at $0.8 < z_{em} < 1.9$, 
and again excluding `associated systems', these authors have found a factor of $\sim$ 2 excess (3$\sigma$ confidence) 
in the incidence of \mgii absorption systems, as compared to the systems 
detected towards QSOs. Interestingly, the excess is found both for strong 
($w_r \ge 1.0\AA$) and weaker ($0.3 \le w_r < 1.0$) \mgii systems, where $w_r$ 
is the rest-frame equivalent width. Thus, the results pertaining to both GRBs 
and BL Lacs hint at a radical premise that the observed occurrence of at 
least the (purportedly intervening) strong Mg II absorption systems
is somehow { connected with} the background source (BBM).

{ 
\begin{table*}
 \centering
 \begin{minipage}[10]{140mm}
\caption{Basic data on our sets of radio-loud quasars (non-blazar type FSRQs).}
\label{tab:basicinfo}
{\scriptsize
\begin{tabular}{@{}lll lll l @{}} 
\hline \hline

  & Data set		  &       Content           & $z_{em}$-range   &  Threshold EW        & Instrument(s) used		& Limiting optical                                   \\           
  & 			  & 			     &  	   &(\mgiia)     & for spectroscopy          &  magnitude                                 \\ 

\hline
                                                                                
1.& Ellison et         &   75 radio-selected      & 0.6$-$1.7            & 0.3\AA	      &  ISIS(WHT 4m)		& unspecified	   	                             \\
  & al. (2004)                 &   quasars (taken\footnote{Applying the 
four selection filters; (i) eliminating the sources that are common to two
or more of the  samples, (ii) excluding all sources that are not in the 
CRATES catalog of flat-spectrum radio quasars (FSRQs), (iii) eliminating
sources that are either classified as `HP' or `BL' in V{\'e}ron-Cetty \& 
V{\'e}ron (2010) or as `BZB' (implying BL Lac), or `BZU' (implying uncertain 
type) in the blazar catalog ROMA-BZCAT (2009) and (iv) availability of a 
intermediate or high-resolution optical spectrum (see Sect.\ 2).}  63)   &    	     &                  & EFOSC2(ESO 3.6m)	&	   	                             \\
  &			  &    			     &		   &    	      &	FORS1 (VLT UT3)         &                                            \\
  &			  &    			     &		   &    	      &	(Intermediate-resolution)         &                                            \\

2.&Bernet et 	          & 77 radio-selected	     & 0.6$-$2.0   &  0.1\AA	      & UVES (VLT UT2)		& $m_{V}<19$	                             \\
  &al. (2010)       	  & quasars (taken$^a$ 32)\footnote{Excluding the 3 quasars   
			  (J114608.1$-$244733, J204719.7$-$163906, 
J213638.6$+$004154) already covered in dataset-1.}
                                   & 		   &  		      & (High-resolution)	    	 	& 		                             \\ 

3.&Jorgenson et          & 68 radio selected       & 0.65$-$1.33   & 0.02\AA	      & 	HIRES (Keck)	& $m_{R}<22.0$   	                     \\
  &al. (2006)	   		   & quasars ({ taken$^a$  8})\footnote{The
68 radio-selected quasars in Jorgenson et al. (2006) were used in the survey 
for damped Ly-$\alpha$ systems. Of these 54 satisfy our selection criteria
and for these we made a search for high resolution spectra in the UVES/VLT 
and HIRES/Keck archives and found reduced HIRES/Keck spectra for 8 of them.}
                                                   &            &           &	(High-resolution)			& (mostly $\sim$18)      \\ 

4.&Narayanan et           & 81 QSOs from ESO	     &0.4$-$2.4    & 0.02\AA	      & UVES (VLT UT2)		& $m_{g}<21.6$   	                     \\ 
  &al. (2007)	     	  & archive (taken$^a$ 12)\footnote{Excluding the 5 
quasars, of which two (J013857.4$-$225447, J204719.7$-$163906) are already 
covered in the dataset-1, and another 3 (J095456.8$+$174331 J113007.0$-$144927 
\& J123200.0$-$022405) in the dataset-2.}            &              &	   & (High-resolution)			 & (mostly $<$19)                           \\

\hline   
\end{tabular}
}                                                            
\end{minipage}                                                           
\end{table*} 
}

Origin of the above unexpected result is unclear. BBM have argued that while 
dust extinction can lower the apparent incidence of absorbers towards QSOs 
and gravitational lensing can increase it towards GRBs and blazars, the 
expected amplitude of these effects falls short, by at least an order 
of magnitude, of explaining the afore-mentioned factor of 2 discrepancy 
between the incidence rates of \mgii absorbers found towards blazars/GRBs 
{ versus } normal QSOs.
They have further estimated that powerful jet in the blazars are capable of 
sweeping sufficiently large column densities of gas (upto $10^{18} - 10^{20} 
cm^{-2}$) and accelerating such clouds to velocities of order $0.1c$, thereby 
possibly accounting for the excess of \mgii absorption systems observed in 
comparison to QSOs. From the observational side, evidence is still lacking 
for such high-velocity outflows of cool material from radio-loud AGN,
although mildly relativistic outflows of highly ionised gas in the nuclear 
region now appear to be commonly  seen for such sources (e.g., Holt et al. 
2008). Based on the observations of Fe XXV/XXVI K-shell resonance lines in 
the X-ray band, outflow speeds of $\approx$ 0.15c have recently been inferred 
for highly ionised gas clouds with column densities of $N_H \approx 10^{23}$ 
cm$^{-2}$ located within the central  parsec of the AGN (Tombesi et al. 2011).
These authors have argued why clouds with even higher { ejection } 
velocities could have remained undetected on account of selection effects 
and instrumental sensitivity limitations. 
However independent evidence  of such high-velocity outflows of cool 
material are still essential  in view of the recent finding of 
Giustini et al. (2011) who have detected highly ionized fast 
($v_x \sim 16 500$\kms) X-ray outflow associated with slower  
UV outflow ($v_x \sim 5000$\kms).

Whilst the above findings lend some credence to the hypothesis that the 
observed excess of \mgii absorption systems towards blazars might have
a causal relationship to gaseous outflows triggered by their relativistic jets
pointed close to our direction (BBM), it would clearly be desirable to 
probe this suggestion using an independent sample of AGN having
powerful Doppler boosted jets. Here we examine this question by
{ using} such a sample, but which consists of a different class of 
powerful AGN, namely flat-spectrum radio quasars (FSRQs, with low optical 
polarization, hence non-blazar type). Compared to the blazar sample 
studied by BBM, our sample of 115 (non-blazar) FSRQs is 2.5 times 
larger and similar or higher resolution optical/UV spectra are available 
for all its members, either from the literature or various data archives. 
\begin{figure}
\epsfig{figure=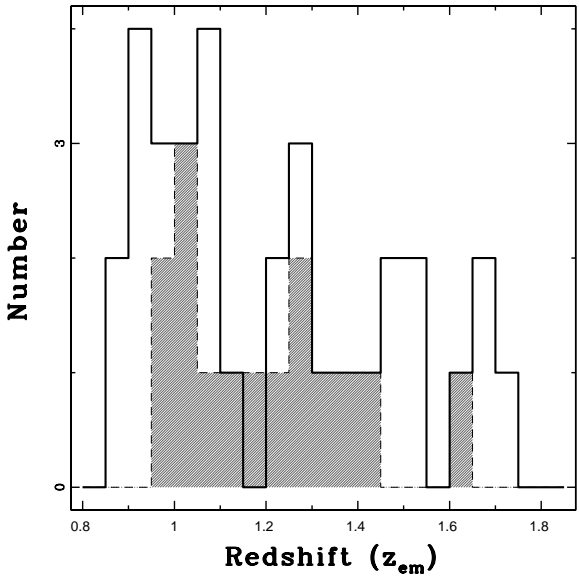,height=8.cm,width=8.cm,angle=00}
\caption[]{{Histograms in emission redshift; solid line is for the 33 
BBM blazars with detected  \mgii absorbing systems (their Table 2).
Shaded region is for our subset of { 15 FSRQs fulfilling the constraints
that (i) emission redshift, $z_{em}$, falls within the range $0.875 
< z_{em} < 1.715$ of the 33 BBM blazars, and (ii)
the detected \mgii absorbing system(s) have $z_{abs}$ in the range 
applicable to the BBM blazars, namely, $0.350 < z_{abs} < 1.430$ for weak 
systems and $0.350 < z_{abs} < 1.579$ for strong absorption systems}.
The K-S test shows that with { $84.8\%$ } probability the two
histograms are drawn from the same parent distribution of $z_{em}$.}}
\label{fig:zhist}
\end{figure}

\section{The FSRQ sample and the spectral data}

{ Our 4 datasets of FSRQs have been derived 
from the four publications listed in Table 1. All but the last of
these publications (Narayanan et al. 2007) provide samples of 
radio-selected quasars; the fourth publication is a mix of radio and 
optical selection (Table 1). To begin with, we merged the samples 
reported in the four publications, obtaining a list of 301 quasars. 
From this basic list we eliminated the 8 quasars present in more
than one of the four samples. For instance,  if a quasars 
had already occurred in 
the first sample, it was deleted from all the remaining samples, if
present therein. From this reduced list to 293 mostly radio-selected
quasars we then excluded all those not contained in the 
CRATES catalog of flat-spectrum radio quasars (FSRQs), which provides 
for each source the spectral index\footnote{$\alpha$ is defined as: 
$flux \propto (frequency)^{\alpha}$}, $\alpha$ between 1.4 GHz and 4.8 
GHz and contains only sources with  $\alpha > -0.5$ and a flux density 
above 65 mJy at 4.8 GHz (Healey et al. 2007). This selection filter left 
us with 201 FSRQs. From this list we further deleted sources that are
either classified as `HP' or `BL' in the V{\'e}ron-Cetty \& V{\'e}ron (2010)
catalogue,
or as `BZB' (implying BL Lac), or `BZU' (implying uncertain type) in 
the catalog ROMA-BZCAT 2009 (Massaro et al. 2009). 
Lastly, we excluded all sources which are absent from both these 
catalogs, treating their nature to be uncertain. This sequence of 
selection filters left us with 163 FSRQs (non-blazars), i.e., 65, 32, 
54 and 12 FSRQs constituting our dataset-1, 2, 3 and 4, respectively,
drawn from the four parent publications mentioned in Table 1.

For the largest dataset-1 (65 southern FSRQs), the parent publication
(Ellison et al.  2004) already provides for all but two sources
(J043850.5$-$201226/B0436$-$203 \& J165945.0$+$021307/B1654-020) the 
entire information relevant for the present study, such as \mgii absorber 
redshift, rest-frame equivalent width EW(\mgiia) and the redshift path values for 
systems having
EW(\mgiia) above the thresholds $0.3\AA,~0.6\AA~$ and $1.0\AA$~(their 
Table~3). Following an unsuccessful search for the optical spectra of
the two remaining sources in the UVES/VLT archive, both sources were 
excluded from further analysis. For the sources in the remaining three 
datasets (nos. 2 -- 4) similar archival search for high resolution 
optical spectra was
much more crucial, since the requisite information was not available 
in their parent publications (Table 1).

For the southern dataset-2, spectra of all 32 FSRQs were found in 
the UVES/VLT\footnote{http://archive.eso.org/eso/eso$_{-}$archive$_{-}$adp.html}
archive. For the northern dataset-3, a search was made in the 
HIRES/Keck\footnote{https://koa.ipac.caltech.edu/cgi-bin/KOA/nph-KOAlogin}
archive and spectra for 10 out of the total 54 FSRQs were found. However, 
for one of them (J064204$+$675836) only raw science frames were available, 
and calibration file was missing. For another FSRQ (J185230$+$401907),
only two exposures of HIRES/Keck observation were available, resulting 
in low signal-to-noise ratio (SNR $<10$) and hence a much reduced 
spectral coverage.
We thus ended up with 8 FSRQs with useful spectra, out of the total
of 54 FSRQs in the dataset-3. Lastly, spectra for all 12 FSRQs in our 
dataset-4 were found in the UVES/VLT archive. This selection process
left us with 115 FSRQs (non-blazar type) with high-quality optical/UV 
spectra, contributed by 63, 32, 8 and 12 FSRQs out of the dataset-1, 2, 3 
and 4, respectively (Table 1). For post-processing of the extracted spectra of 
the FSRQs belonging to datasets 2--4, such as air to vacuum wavelength 
conversion, heliocentric correction, combining individual exposures to 
enhance the SNR and continuum fitting, we have  followed the procedure 
described in Chand et. al (2004).  Details of our final 
sample\footnote{\it{Full table is available only in on-line version.}} 
of { 115} FSRQs are given in Table~\ref{tab:samp} }.\par

\begin{figure*}
\epsfig{figure=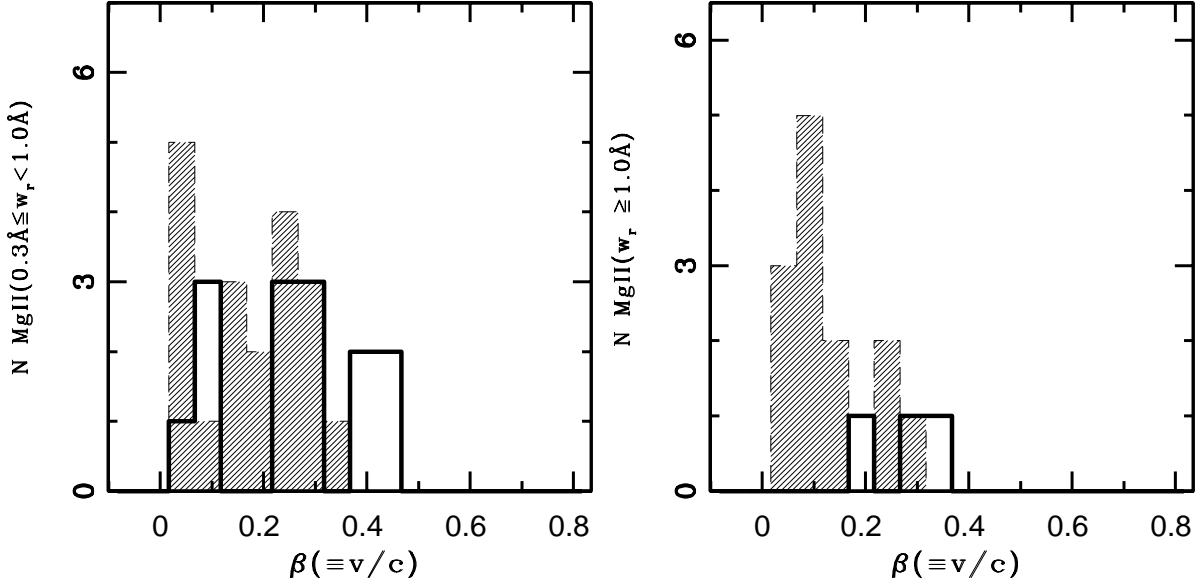,height=8.cm,width=16.cm,angle=00,bbllx=34bp,bblly=160bp,bburx=568bp,bbury=430bp,clip=true}

\caption[]{{ \emph{Left:} Histogram  of the weak $(0.3\AA~\le w_{\rm r}(2796) 
< 1.0 ~\AA~$) `intervening' Mg\,{\sc ii} absorption systems towards the 
FSRQs, w.r.t. the velocity of the absorber relative to the background 
FSRQ (Eq.~\ref{eq:beta}), 
(thick, black curve). The shaded region shows the corresponding histogram 
of the weak Mg\,{\sc ii} absorption systems found towards the 
BBM blazars, adopting the incidence of these systems as given in Bergeron 
et al. (2011).}{{ \emph{Right:} Same as the left panel, but for strong 
Mg\,{\sc ii} absorption systems $( w_{\rm r}(2796) \ge 1.0$~\AA~).}}  
{ This comparison is carried out using a subset of 15 FSRQs drawn
from our FSRQ sample, by applying a redshift matching procedure
described in Sect. 4 and after excluding the systems with $\Delta v 
< 5000$ km.s$^{-1}$, i.e., $\beta<0.017$.}}
\label{fig:beta}
\end{figure*}
{
\begin{table*}[t]
\begin{center}
\begin{minipage}[10]{130mm}
\caption{Comparison of \mgii absorption systems towards FSRQs and blazars.}
\label{tab:sumary}
{\scriptsize
\begin{tabular}{@{}rccc ccccc @{}} 
\hline \hline
\multicolumn{1}{c}{Absorber type}
& \multicolumn{1}{c}{$w_{r}(2796)$-range}
& \multicolumn{1}{c}{N$_{obs}$}
& $\Delta$z\footnote{Redshift path estimated for the specified EW detection
    threshold above 5$\sigma$ significance level.}
& $\langle z \rangle$ 
& $\frac{dN}{dz}$ 
& $\left(\frac{dN/dz}{(dN/dz)_{qso}} \right) $
&$\eta(\langle z \rangle)$ \footnote{The scale factor $\eta(\langle z \rangle)$ is
used for scaling our computed $dN/dz$ for the FSRQ weak and strong absorption
systems,  from their respective mean redshifts (col. 5) to the mean redshift 
of the BBM blazars 
($\langle z \rangle$ = 0.83 for the weak and 0.82 for the strong \mgii absorption 
systems). For details, see the text in the beginning of  Sect.~4.}
&$\left(\frac{\eta(\langle z \rangle)(dN/dz)}{(dN/dz)_{Blz}} \right)$\footnote{$(dN/dz)_{Blz}=0.76_{0.17}^{0.22}~ \mbox{at}~ \langle z \rangle~ =~ 0.83$ for the weak absorption systems and 
$(dN/dz)_{Blz}~=~0.43_{0.12}^{0.16}~\mbox{at}~ \langle z \rangle~ =~0.82$ 
for the strong absorption systems detected towards the BBM blazars (their Eqs. 5 $\&$ 8). } \\
\hline \\

Weak systems  &$~0.3~$\AA$ \le w_{r}(2796)<1.0$ &  53 & 80.23 & $1.12\pm0.42$&$   0.66_{0.09}^{0.10}$&$1.40_{0.19}^{0.22}$&0.91&$   0.79_{0.19}^{0.27}$ \\\\

Strong systems&$~1.0~$\AA$\le w_{r}(2796)$      &  22 &106.02 &$1.09\pm0.42$&$   0.21_{0.04}^{0.05}$&$0.85_{0.18}^{0.22}$&0.81&$   0.39_{0.12}^{0.19}$ \\\\


\hline   
\end{tabular} 
}                                                           
\end{minipage}                                                           
\end{center}
\end{table*} 
}
\section{Analysis}
Although both UVES and HIRES spectrographs provide a large wavelength
coverage, from $\sim$ 3000\AA~ to $\sim$ 10000\AA~, the available 
spectral coverage varies from quasar to quasar, based on the choice of 
cross-disperser settings. Combining exposures from various settings, 
therefore, resulted sometimes in gaps in the wavelength coverage. 
In addition, we systematically excluded the following spectral path 
lengths from our formal search for \mgii absorption systems: 
(i) wavelength region blueward of the Ly-$\alpha$ 
emission line, to avoid contamination  by the Ly-$\alpha$ forest; 
(ii) wavelength regions within 5000~\kms of the \mgiiab emission 
lines at the quasar redshift, as any absorption lines within this
velocity interval have a higher probability of being associated with the 
quasar itself (Sect.\ 1); and (iii) spectral regions polluted by the various
known atmospheric absorption features; these were eliminated by eye 
for any line blending. Further, noisy spectral regions got discarded
by our considering only the spectral ranges with SNR sufficient to 
detect with $> 5\sigma$ significance any absorption line above the 
adopted EW threshold (which is 0.3\AA~ for weak systems and 1.0\AA~ 
for strong systems); this is also the criterion used in Ellison et al. 
2004 (the parent paper for our dataset-1).\par

In searching for {\mgii} absorption systems, within the included 
redshift path in each quasar spectrum, we first plotted the normalized 
spectrum and then plotted over it the same spectrum by shifting the 
wavelength array by a factor of $\lambda2796.3543/\lambda2803.5315$. 
Small spectral segments of about 50~\AA~ were then visually inspected 
for comparable profile shapes and for a doublet ratio between 1:1 and 2:1.
To further ascertain the detection of the absorption system, we then 
searched for corresponding metal lines (e.g. {\feii}, {\mgii}, {\civ}, 
{\siiv}, etc.) in the velocity plot. Weak {\mgii} doublets that were 
found within $500$~{\kms} of each other were taken to be part of the 
same absorption system and therefore classified as one multi-cloud 
system as is { usually done}.

 From the above analysis of the optical spectra of the total { 52} FSRQs 
drawn from the datasets 2--4, we detected 
{ 29} {\mgii} absorption systems with $0.3$ \AA$ \le w_{r}(2796)<1.0$ \AA~ 
(``weak systems''), and { 10} {\mgii} absorption systems with $w_{r}(2796) 
\ge 1.0$ \AA~  (``strong systems''). Thus, including  the dataset-1, 
the total counts of weak and strong \mgii absorption systems in our 
final sample of  115 FSRQs become  53 and 22, respectively 
(Table~4)$^4$. The detected \mgii absorbing system(s) have $z_{abs}$ 
range,  median redshift and redshift dispersion, respectively, of  $~0.399 < z_{abs} < 2.638$,  
$1.119$ and $0.499$ for the weak systems and $0.238 < z_{abs} < 1.969$, $0.828$, $0.442$ 
for the strong absorption systems.
 
Table~\ref{tab:sumary} summarizes the results for our final sample of 115
FSRQs, giving $dN/dz=N_{obs}/\Delta z$, 
where $N_{obs}$ is the number of the observed \mgii absorption systems
within the redshift path $\Delta z$; 
for both strong and weak \mgii systems.
The errors on $dN/dz$ are calculated based on Poisson statistics for small 
numbers, with the limits corresponding to 1$\sigma$ confidence level of a 
Gaussian distribution, as tabulated by Gehrels (1986).

\section{Summary and Discussion}

 It is interesting to note from Table~\ref{tab:sumary} that the 
$dN/dz$ for FSRQs and normal QSOs is quite similar both for weak and 
strong \mgii absorption systems (e.g see col. 7 Table~\ref{tab:sumary}). Here, to 
compute $dN/dz$ for normal QSOs at the mean redshift of our FSRQs sample, 
we use the fit of $z$ versus $dN/dz$ for strong 
absorption system ($w_r(2796)\ge 1.0\AA$), as given by Prochter et al. (2006b)
(see also Eq. 6 of BBM)  and for $w_r(2796)\ge 0.3\AA$~
systems as given by  Nestor (e.g., see Eq.~2 of BBM, their pvt. comm.), 
derived in both cases by using the QSOs in the SDSS (DR4). Thus, for weak 
systems toward QSOs, $dN/dz$ at any redshift can be computed by taking 
the difference between the above two fits (i.e., $dN/dz$ for
$w_r(2796)\ge 0.3\AA$~ minus $dN/dz$ for 
$w_r(2796)\ge 1.0\AA$). However, for carrying out the comparison with the 
result reported by BBM for blazars (e.g, see col. 9 of Table~\ref{tab:sumary} ), 
which has different mean redshift compare to mean redshift of our 
FSRQs sample (e.g, see col. 5  of Table~\ref{tab:sumary}), 
we have to first take into account the known cosmological evolution of 
$dN/dz$, by scaling our $dN/dz$ values found for the FSRQs, to the mean
redshift of the \mgii absorption systems towards the BBM blazars 
($\langle z \rangle~ =~ 0.83$). To do this scaling, we have 
assumed that the cosmological evolution of $dN/dz$ for our FSRQ sample 
is similar to that known for $dN/dz$ towards QSOs, which seem reasonable 
 considering (i)  the  lack of  $dN/dz$ 
evolution studies for radio-loud quasars and (ii) that our $dN/dz$ 
estimated for FSRQs appears quite similar to that known for 
QSOs (see above). The computed scale factors, $\eta(\langle z \rangle)$ 
are listed in col. 8 of Table~\ref{tab:sumary}. 
By multiplying with these scale factors, our computed values of $dN/dz$ 
for the FSRQ subsets are scaled to the (slightly lower) mean redshift of 
the BBM blazar sample ($\langle z \rangle~ =~ 0.83$), before the ratio 
of $dN/dz$ for our FSRQ subsets and the corresponding BBM blazar sets 
are taken ( col. 9 in Table 2).

The most interesting result from Table~\ref{tab:sumary} is that for 
 \mgii strong absorption systems, $dN/dz$ for FSRQs is only about half of the 
value reported by BBM for blazars, which they showed to be itself nearly 
twice the value known for optically selected quasars (QSOs, which are mostly radio quiet).
In other words, the excess abundance of \mgii strong 
absorption systems seen towards blazars (and also GRBs), in comparison to
QSOs (BBM and references therein), does not seem characteristic of 
flat-spectrum radio quasars of non-blazar type, even though they too 
possess powerful Doppler boosted jets. 
BBM have suggested that the excess seen towards blazars might be due 
to absorbing gas clouds swept up by the powerful jets and accelerated 
to mildly relativistic velocities (Sect.\ 1). While this might readily
explain the excess of the absorption systems relative to normal quasars
(i.e., optically selected, mostly radio-quiet, QSOs) which lack powerful 
jets, can this be reconciled with the lack of excess found here for FSRQs 
which do possess powerful relativistic jets? A relevant factor here is 
the likelihood that, compared to blazars (and GRBs), the jets in 
FSRQs may be less well aligned to the line of sight.
{ A plausible outcome of such a jet orientation scenario, which 
underlies a popular class of unification models of radio-loud AGN 
( { e.g., Orr \& Browne 1982; Antonucci \& Ulvestad 1985; Wills \& 
Browne 1986}), would be that any gas clouds accelerated by the powerful FSRQ 
jets to extremely high (even mildly relativistic) speeds may simply
not appear in the foreground of the quasar nucleus and hence escape 
detection as absorption systems.}

To probe this hypothesis, we now compare for the FSRQs and blazars the 
distributions of speeds, $\beta$, of the observed \mgii absorption-line 
systems, relative to the parent quasar/blazar, where

\begin{equation}
\beta \equiv \frac{v}{\rm c}  = \frac {(1+z_{\rm em})^2-(1+z_{\rm abs})^2}
{(1+z_{\rm em})^2+(1+z_{\rm abs})^2}.  
\label{eq:beta}
\end{equation}

To make a meaningful comparison we have derived a subset from our FSRQ 
sample, by taking only those { 15 FSRQs which satisfy the constraints that 
(i) the emission redshift, $z_{em}$, falls within the range $0.875 < z_{em}
< 1.715$ of the  33 BBM blazars actually showing \mgii absorption systems 
(out of the total 45 blazars listed in Table 2 of BBM)), and (ii) the
detected \mgii absorption systems satisfy $0.350 <z_{abs} < 1.430$ 
(for weak systems) and $0.350 <z_{abs} < 1.579$ (for strong systems), being 
the $z_{abs}$ ranges for the 33 BBM blazars}.
To begin with, histograms of $z_{em}$ for the subset of these { 15} FSRQs 
and the 33 BBM blazars are compared in Fig.~\ref{fig:zhist}. K-S test shows that with 
{ $84.8\%$} probability the two histograms are drawn from the same intrinsic 
distribution of $z_{em}$. In view of the known cosmological evolution of 
$dN/dz$, this prior confirmation of redshift matching is important.
{ After the above matching of the redshift range both in 
$z_{em}$ and $z_{abs}$, which led to our subset of 15 FSRQs for comparison 
with the 33 BBM blazars, we are left with 14 weak and 3 strong \mgii
absorption systems seen towards our 15 FSRQs, to be compared with the 19 
weak and 13 strong \mgii absorption systems 
detected towards the 33 BBM blazars.}
The two panels in Fig.~\ref{fig:beta} compare $\beta$ distributions 
for the redshift matched subsets of our FSRQs and BBM blazars, separately 
for the weak and strong Mg II absorption systems (recall that `associated 
systems' having $\beta<$ 0.017 have already been excluded, Sect. 3). 
 The unpaired t-test  probabilities that the $\beta$ distributions for 
blazars and FSRQs are drawn from the same parent population, are $99.99\%$
and $1.3\%$ for the weak and strong absorption systems, respectively. 
Here unpaired t-test is preferred over the KS-test, in view of the small 
sample size of the absorption systems, especially the strong systems. 
Thus, the difference between the strong absorption systems towards blazars and 
FSRQs appears statistically significant, although larger datasets would 
clearly be very desirable. As seen from Fig. 2 (right panel), the main 
contributor to the difference between the two $\beta-$distributions is the
conspicuous presence of strong absorption systems for the BBM blazars at
$\beta$ values up to 0.15. Interestingly, BBM have reported a similar 
excess of strong \mgii absorption systems towards their blazars, in 
comparison to QSOs (their Fig. 4), which they interpret by postulating 
high-speed outflowing clouds of cool gas accelerated by the powerful blazar 
jets (Sect.\ 1). While the FSRQs studied here also possess powerful jets, 
the deficit of strong absorption systems at lower $\beta$ is still 
present, in comparison to blazars (Fig.\ 2 right panel). As noted above, 
a plausible way to reconcile this with the BBM hypothesis is then to 
consider the possibility that the FSRQ jets are less closely aligned to 
our direction, compared to the jets in the BBM blazars which are also powerful 
and expected (by definition) to be strongly polarized. Such an 
inference was indeed drawn by Valtaoja et al. (1992), based purely on 
their extensive radio flux monitoring data at centimeter wavelengths, 
which showed that, { on average}, low-polarization quasars (FSRQs) 
vary with a smaller amplitude and on longer time scale, compared to blazars 
(see, also, Orr \& Browne 1982). One would then expect that any ultra-fast 
moving absorbing clouds accelerated by powerful FSRQ jets would mostly 
be out of the line of sight to the quasar nucleus. This suggestion can be tested by the analysis of 
\mgii absorption systems towards {\it steep-spectrum} quasars 
whose jets are expected to be even 
less well aligned to our direction (e.g., Barthel 1999; Orr \& Browne 1982). 

Another hint favoring such an orientation based explanation comes from 
the radio properties 
of the small minority of FSRQs that have been detected with the Large Area 
Telescope (LAT) on board the {\it Fermi Gamma-ray Space Telescope}(see Linford 
et al. 2011).  These authors find that the $\gamma-$ray bright LAT FSRQs are 
extreme sources with higher core brightness temperature and greater core 
polarization, as well as larger (apparent) opening angle of the parsec-scale 
jets, than their non-LAT counterparts. Interpreting these differences, they 
have suggested that the $\gamma-$ray loud FSRQs can be explained by Doppler 
boosting, but the jet orientation and/or speeds must be significantly 
different than for the $\gamma-$ray quiet FSRQs (which show weaker 
polarization). This scenario would be consistent with the above 
{ assumption} that the jets in our sample of (weakly polarized, hence 
non-blazar) FSRQs may be less closely aligned to the line of sight, in 
comparison to the similarly powerful jets in the BBM blazars.  Although,
such a bias may well explain the difference between the occurrence rates
of strong \mgii absorption towards powerful blazars and FSRQs/QSOs, further
substantiation of this trend employing larger datasets would greatly
strengthen the BBM scenario that strong \mgii absorption systems may often 
be revealing the most extreme velocity outflows of cool gas clouds from 
powerful AGN, underscoring the need to investigate the wide-ranging 
theoretical implications of this phenomenon.
  

\acknowledgments
This research has made use of (i) the Keck Observatory Archive (KOA),
which is operated by the W. M. Keck Observatory and the NASA Exoplanet
Science Institute (NExScI), under contract with the National
Aeronautics and Space Administration, by using observations done using
HIRES spectrograph at the Keck, Mauna Kea, Hawaii (ii) ESO Science
Archive Facility by using observation done using the UVES spectrograph at
the VLT, Paranal, Chile and (iii)  the NASA/IPAC Extragalactic Database (NED) 
which is operated by the Jet Propulsion Laboratory, California Institute of 
Technology, under contract with the National Aeronautics and Space Administration.\\

We thank  an anonymous referee for the constructive criticism and helpful suggestions.
We gratefully acknowledge the help from T. Parasakthi with the processed 
data for the quasars belonging to our dataset-4, and from Ravi Joshi in 
resolving some issues related to the archival data used here.

{\it Facilities:} \facility{Keck Observatory Archive \& ESO Science
Archive Facility}.

\begin{table*}
\begin{center}
\caption{Basic properties of our sample of 115 (non-blazar) FSRQs}
\label{tab:samp}
{\scriptsize
\begin{tabular}{@{}lrc c rcrl ccl@{}}
\hline \hline
\multicolumn{1}{c}{Quasar name}  
& \multicolumn{1}{c}{z$_{em}$$^a$} 
& \multicolumn{1}{c}{$\alpha_{radio}$} 
& \multicolumn{1}{c}{f(4.8GHz)}
& \multicolumn{1}{c}{mag$^b$} 
& \multicolumn{1}{c}{filter$^b$} 
& \multicolumn{2}{c}{Weak systems} 
& \multicolumn{2}{c}{Strong systems}
&\multicolumn{1}{l}{Ref. code}\\ 
 \multicolumn{1}{c}{}
& \multicolumn{1}{c}{} 
& \multicolumn{1}{c}{} 
& \multicolumn{1}{c}{(mJy)}
& \multicolumn{1}{c}{} 
& \multicolumn{1}{c}{} 
& \multicolumn{1}{c}{$\Delta z$}
& \multicolumn{1}{c}{$N_{sys}$}
& \multicolumn{1}{c}{$\Delta z$}
& \multicolumn{1}{c}{$N_{sys}$}
& \multicolumn{1}{l}{(see Table 1)}\\
 \hline  
J001130.5$+$005551  &      2.308620 &      -0.142 &    140  & 19.10 & V &   1.1656 &  0   &     1.1656 &  0& 4 \\
J001602.4$-$001225  &      2.086940 &      -0.443 &    645  & 18.36 & V &   1.0823 &  0   &     1.0823 &  0& 4 \\
J001708.5$+$813508  &      3.366000 &      -0.200 &    551  & 16.52 & V &   0.5544 &  0   &     0.5544 &  0& 3 \\
J004057.6$-$014632  &      1.178000 &      -0.017 &    581  & 18.30 & O &   0.9942 &  1   &     0.9942 &  0& 2 \\
J004201.2$-$403039  & 	   2.478000&       0.336 &    299  & 19.93 & P &   0.0114 &  0   &	0.7135 &  1& 1 \\
  ..........        &        .....  &       ...   &   ...   &  .....& . &  ......  &  .   &     .....  &  .&  . \\   
\hline
\multicolumn{11}{l}{\it{$^{a}$ Based on  NASA/IPAC Extragalactic Database (NED).}}\\ 
\multicolumn{11}{l}{\it{$^{b}$ Based on  V{\'e}ron-Cetty \& V{\'e}ron(2010).}}\\ 
\multicolumn{11}{l}{{ Note.} The entire table for our sample of  { 115} FSRQs is available 
in on-line version.}\\
\multicolumn{11}{l}{Only a portion of this table is shown here to display its form and content.}\\ 

\end{tabular}
}
\end{center}
\end{table*}


\begin{table*}
\begin{center}
\caption{\mgii absorption systems and their rest-frame  
equivalent width , $w_{r}$(Mg{\sc ii}$\lambda 2796\AA$), for our sample of  115 (non-blazar) FSRQs.}
\label{tab:sampabs}
\begin{tabular}{@{}lrc rr ccl@{}}
\hline \hline
\multicolumn{1}{c}{FSRQ}  
& \multicolumn{1}{c}{z$_{em}$} 
& \multicolumn{1}{c}{z$_{abs}$} 
& \multicolumn{1}{c}{w$_{r}$(2796\AA)} 
& \multicolumn{1}{c}{$\alpha_{radio}$} 
& \multicolumn{1}{c}{f(4.8GHz)}
& \multicolumn{1}{c}{$\sim$SNR}
&\multicolumn{1}{l}{Ref. code}\\ 
  \multicolumn{1}{c}{}
& \multicolumn{1}{c}{}
& \multicolumn{1}{c}{}
& \multicolumn{1}{c}{(\AA)}
& \multicolumn{1}{c}{} 
& \multicolumn{1}{c}{(mJy)}
& \multicolumn{1}{l}{per FWHM}
& \multicolumn{1}{l}{(see Table 1)}\\
 \hline  
J004057.6-014632 &  1.1780 &  0.6828 &   $0.36  \pm  0.01$ & $-$0.02   &    581.0   &    22.2  &   2 \\
J004201.2-403039 &  2.4780 &  0.8483 &   $2.35  \pm  0.15$ & $+$0.34   &    299.0   &   .....  &   1 \\
J005108.2-065002 &  1.9750 &  1.4919 &   $0.90  \pm  0.03$ & $-$0.08   &    841.0   &   .....  &   1 \\
J005108.2-065002 &  1.9750 &  1.5698 &   $0.32  \pm  0.03$ & $-$0.08   &    841.0   &   .....  &   1 \\
J024008.2-230916 &  2.2230 &  1.3652 &   $1.86  \pm  0.01$ & $-$0.46   &   3630.0   &   261.1  &   4 \\
  ..........     &  .....  &  ...... &   $....  \pm  ....$ & $ $....   &   ......   &   .....  &   . \\   
\hline
\multicolumn{8}{l}{{ Note.} The entire table of our total 75 \mgii absorbers is available in on-line version.}\\
\multicolumn{8}{l}{Only a portion of this table is shown here to display its form and content.}\\ 

\end{tabular}
\end{center}
\end{table*}

\end{document}